\documentclass[sigconf]{acmart}
\AtBeginDocument{%
  }

\copyrightyear{2026}
\acmYear{2026}
\setcopyright{cc}
\setcctype{by}
\acmConference[KDD '26]{Proceedings of the 32nd ACM SIGKDD Conference on Knowledge Discovery and Data Mining V.2}{August 09--13, 2026}{Jeju Island, Republic of Korea}
\acmBooktitle{Proceedings of the 32nd ACM SIGKDD Conference on Knowledge Discovery and Data Mining V.2 (KDD '26), August 09--13, 2026, Jeju Island, Republic of Korea}
\acmDOI{10.1145/3770855.3817575}
\acmISBN{979-8-4007-2259-2/2026/08}

\usepackage{latexsym}
\usepackage{enumitem}

\usepackage{graphicx}
\usepackage{multirow}
\usepackage{booktabs}
\usepackage{subcaption}
\usepackage[most]{tcolorbox}

\usepackage{float}
\usepackage{placeins}

\begin{document}

\title{RAG vs. GraphRAG: A Systematic Evaluation and Key Insights}


\author{Haoyu Han}
\affiliation{
  \institution{Michigan State University}
  \city{East Lansing}
  \state{MI}
  \country{USA}
}

\author{Li Ma}
\authornote{Corresponding Authors.}
\affiliation{
  \institution{Michigan State University}
  \city{East Lansing}
  \state{MI}
  \country{USA}
}

\author{Yu Wang}
\affiliation{
  \institution{University of Oregon}
  \city{Eugene}
  \state{OR}
  \country{USA}
}

\author{Harry Shomer}
\affiliation{
  \institution{University of Texas at Arlington}
  \city{Arlington}
  \state{TX}
  \country{USA}
}

\author{Kai Guo}
\affiliation{
  \institution{Michigan State University}
  \city{East Lansing}
  \state{MI}
  \country{USA}
}

\author{Yongjia Lei}
\affiliation{
  \institution{University of Oregon}
  \city{Eugene}
  \state{OR}
  \country{USA}
}

\author{Zhisheng Qi}
\affiliation{
  \institution{University of Oregon}
  \city{Eugene}
  \state{OR}
  \country{USA}
}

\author{Zhigang Hua}
\affiliation{
  \institution{Meta Platforms, Inc.}
  \city{Menlo Park}
  \state{CA}
  \country{USA}
}

\author{Bo Long}
\affiliation{
  \institution{Meta Platforms, Inc.}
  \city{Menlo Park}
  \state{CA}
  \country{USA}
}

\author{Hui Liu}
\affiliation{
  \institution{Michigan State University}
  \city{East Lansing}
  \state{MI}
  \country{USA}
}

\author{Charu Aggarwal}
\affiliation{
  \institution{IBM T.J. Watson Research Center}
  \city{Yorktown Heights}
  \state{NY}
  \country{USA}
}

\author{Jiliang Tang}
\affiliation{
  \institution{Michigan State University}
  \city{East Lansing}
  \state{MI}
  \country{USA}
}








\renewcommand{\shortauthors}{Haoyu Han et al.}

\begin{abstract}
Retrieval-Augmented Generation (RAG) improves large language models (LLMs) by retrieving relevant information from external sources and has been widely adopted for text-based tasks. For structured data, such as knowledge graphs, Graph Retrieval-Augmented Generation (GraphRAG) retrieves and aggregates information along graph structures. More recently, GraphRAG has been extended to general text settings by organizing unstructured text into graph representations, showing promise for reasoning and grounding.
Despite these advances, existing GraphRAG systems for text data are often tailored to specific tasks, datasets, and system designs, resulting in heterogeneous evaluation protocols. Consequently, a systematic understanding of the relative strengths, limitations, and trade-offs between RAG and GraphRAG on widely used text benchmarks remains limited.
In this paper, we present a comprehensive benchmark study comparing RAG and GraphRAG on established text-based tasks, including question answering and query-based summarization. We introduce a unified evaluation protocol that standardizes data preprocessing, retrieval configurations, and generation settings, enabling fair and reproducible comparisons. Our results highlight the distinct strengths of RAG and GraphRAG across different tasks and evaluation perspectives. 
Building on these findings, we explore selection and integration strategies that combine the strengths of both paradigms, leading to consistent performance improvements. We further analyze failure modes, efficiency trade-offs, and evaluation biases, and highlight key considerations for designing and evaluating retrieval-augmented generation systems. Our code is available at https://github.com/haoyuhan1/RAGvsGraphRAG.

\end{abstract}





\maketitle

\section{Introduction}
\label{sec:intro}
Retrieval-Augmented Generation (RAG) has emerged as a practical paradigm for improving downstream tasks by retrieving relevant knowledge from external data sources. It has been successfully deployed in a wide range of real-world applications, including healthcare~\citep{xu2024ram}, law~\citep{wiratunga2024cbr}, finance~\citep{zhang2023enhancing}, and education~\citep{miladi2024leveraging}. With the advent of Large Language Models (LLMs), integrating retrieval into generation further improves faithfulness by mitigating hallucinations and enhancing robustness~\cite{zhao2023survey, huang2023survey}. In most existing RAG systems, retrieval is performed over text corpora.

Graphs provide an explicit representation of relational structure and have long been used across domains such as knowledge representation, social networks, and biomedical discovery~\cite{wu2020comprehensive, ma2021deep, wu2023survey}. Graph Retrieval-Augmented Generation (GraphRAG) has recently gained attention for retrieving and aggregating information from graph-structured data, including knowledge graphs (KGs) and molecular graphs~\citep{han2024retrieval, peng2024graph}. Beyond leveraging existing graphs, an emerging line of work extends GraphRAG to text-based tasks by constructing graphs from unstructured documents, with reported benefits for global summarization~\cite{edge2024local}, planning~\cite{lin2024graph}, and reasoning~\cite{han2025reasoning}. However, most GraphRAG studies for text are conducted under task- and system-specific settings, using bespoke datasets, graph construction heuristics, and evaluation protocols. This heterogeneity makes it difficult to draw principled conclusions about \textit{when and why} explicit graph structures help (or hurt) retrieval-augmented generation, and obscures practical trade-offs such as construction cost, retrieval latency, and storage footprint.

To address this gap, we conduct a controlled and systematic benchmark of RAG and GraphRAG on widely used text-based tasks, focusing on \textit{Question Answering} (QA) and \textit{Query-based Summarization}. We consider four representative categories of GraphRAG systems: 
{\bf (1)} \emph{KG-based GraphRAG}~\cite{Liu_LlamaIndex_2022}, which extracts a KG from text and performs retrieval over the KG; 
{\bf (2)} \emph{Community-based GraphRAG}~\citep{edge2024local}, which performs retrieval over community structures and hierarchical abstractions; 
{\bf (3)} \emph{Text-centric graph-guided RAG}~\citep{jimenez2024hipporag}, which retrieves original text chunks with the assistance of a constructed knowledge graph; and 
{\bf (4)} \emph{Hierarchical summary-based GraphRAG}~\citep{sarthi2024raptor}, which builds hierarchical summaries to enable multi-granular retrieval without relying on explicit KGs.
For QA, we evaluate both single-hop and multi-hop settings; for summarization, we consider both single-document and multi-document scenarios. Crucially, we introduce a unified evaluation protocol that standardizes data preprocessing, retrieval, and generation settings, enabling fair and reproducible comparisons across paradigms.

Our analysis results lead to several key findings. First, \textbf{RAG and GraphRAG exhibit complementary behaviors rather than a consistent winner}. In QA, RAG performs better on single-hop and detail-oriented factual queries, whereas GraphRAG is more effective on multi-hop, reasoning-intensive questions. Second, \textbf{GraphRAG design choices matter}: for example, community-based global search can sacrifice query-specific details, hurting detail-oriented QA, while providing more corpus-level aggregation that benefits broad or diverse summarization outputs. Third, \textbf{evaluation protocol can change conclusions}: we show that LLM-as-a-Judge evaluation for summarization can be highly sensitive to the presentation order of candidate summaries, introducing strong position effects that may confound comparisons. Finally, \textbf{GraphRAG is not free}: it often incurs higher construction cost, retrieval latency, and storage footprint, and its performance can be sensitive to the quality (and cost) of graph construction.
These findings suggest that effective retrieval-augmented generation should not treat RAG and GraphRAG as mutually exclusive choices. Motivated by this, we study two practical hybrid strategies: \textbf{Selection}, which routes queries to RAG or GraphRAG based on query type for efficiency, and \textbf{Integration}, which combines evidence from both paradigms to maximize performance. Across benchmarks, these strategies yield consistent improvements.
Our main contributions are as follows:
\begin{itemize}[leftmargin=*]
    \item \textbf{Systematic Benchmark:} We present a controlled benchmark comparing RAG and multiple GraphRAG variants across QA and query-based summarization under a unified evaluation protocol (consistent preprocessing, retrieval, and generation settings) for fair and reproducible comparison.
    \item \textbf{Strong, Task-Level Findings:} We identify clear complementarities: RAG is stronger for factual/detail-oriented QA, while GraphRAG benefits reasoning-intensive QA and produces more corpus-level, diverse summaries, with outcomes strongly affected by GraphRAG design choices (e.g., local vs.\ global search).
    \item \textbf{Hybrid Strategies:} We study \textbf{Selection} and \textbf{Integration} strategies that combine RAG and GraphRAG, achieving consistent improvements and illustrating effectiveness--efficiency trade-offs.
    \item \textbf{Evaluation and Efficiency Analyses:} We analyze failure modes, construction/retrieval/storage costs, sensitivity to graph construction quality, and demonstrate strong position effects in LLM-as-a-Judge summarization evaluation, highlighting practical considerations for reliable (Graph)RAG assessment.
\end{itemize}

\vspace{-0.1in}
\section{Related Works}
\label{sec:related}
Retrieval-Augmented Generation (RAG) has been widely used to enhance LLMs by retrieving relevant information from external sources~\cite{fan2024survey, gao2023retrieval}. 
Most RAG systems process text corpora by splitting documents into passages or chunks~\cite{finardi2024chronicles}. 
Given a query, relevant chunks are retrieved using lexical search~\cite{ram2023context}, dense semantic retrieval~\cite{karpukhin2020dense}, or hybrid retrieval strategies. 
Beyond vanilla retrieval, prior work has explored pre-retrieval processing, such as query rewriting and query expansion~\cite{ma2023query, zheng2023take}; post-retrieval processing, such as reranking, filtering, and compression~\cite{dong2024don, xu2023recomp}; and fine-tuning strategies to better align retrievers or generators with downstream tasks~\cite{li2023structure}. 
These techniques have improved RAG across a wide range of applications, including question answering~\cite{yan2024corrective}, dialogue generation~\cite{izacard2023atlas}, and summarization~\cite{jiang2023active}. 
Recent studies also benchmark RAG pipelines and evaluation tools across tasks, domains, and metrics~\cite{yu2024evaluation, chen2024benchmarking, es2023ragas}. 

While standard RAG typically treats documents as independent text chunks, many real-world information sources contain rich relational structures, such as knowledge graphs, social graphs, citation networks, and molecular graphs~\cite{xia2021graph, ma2021deep}. 
Graph Retrieval-Augmented Generation (GraphRAG) incorporates such graph structures into retrieval and generation to exploit relational signals among connected entities, documents, or chunks~\cite{han2024retrieval, peng2024graph}. 
Early work mainly focuses on retrieval over existing knowledge graphs for downstream tasks such as knowledge-graph question answering~\cite{tian2024graph, yasunaga2021qa} and fact checking~\cite{kim2023factkg}. 
Graph structures have also been used to improve text-centric retrieval; for example, hyperlink or document graphs can guide evidence selection for question answering~\cite{li2022dynamic}.
More recently, GraphRAG methods have begun to construct graphs directly from text corpora to support general text-based tasks~\cite{han2024retrieval}. 
One line of work builds document- or chunk-level graphs to guide retrieval over textual units~\cite{dong2024don, li2022dynamic}. 
Another line constructs entity--relation graphs, often with LLM assistance, and retrieves information at multiple abstraction levels, such as local entity neighborhoods or global community-level summaries~\cite{edge2024local, han2025reasoning}. 
Related hierarchical approaches use graph-inspired structures without requiring fully specified entity--relation semantics. For example, RAPTOR constructs hierarchical summary trees for multi-granular retrieval~\cite{sarthi2024raptor}, while HippoRAG and its extensions build entity-linked memory graphs to guide chunk retrieval~\cite{jimenez2024hipporag, gutierrez2025rag}.

Despite rapid progress, GraphRAG systems for text are often evaluated under heterogeneous protocols, with varying graph construction methods, retrieval configurations, and even evaluation criteria. 
Moreover, graph construction introduces additional costs (indexing time, retrieval latency, and storage footprint) and can be sensitive to the quality of the construction model, yet these trade-offs are not consistently characterized across studies. 
As a result, it remains unclear how GraphRAG compares with standard RAG on general text-based benchmarks and what practical trade-offs it entails. This motivates our systematic benchmark evaluation under unified experimental settings.

\section{Evaluation Framework}
\label{sec:background}
In this section, we describe our evaluation framework
 and experimental protocol. To ensure fair comparison, we evaluate RAG and GraphRAG under identical settings whenever applicable, and otherwise follow the default configurations of each method while matching key budgets. We decouple retrieval from generation by first saving retrieved evidence for each method and then using a unified generation script to produce outputs conditioned on the saved retrieval results.

\subsection{RAG Pipeline}
We adopt a standard dense-retrieval RAG pipeline~\cite{karpukhin2020dense}. 
Given a corpus, we segment documents into textual chunks and build an index by embedding each chunk into a shared vector space. 
At inference time, we embed the query, retrieve top-ranked chunks based on the semantic similarity.

\subsection{GraphRAG Implementations}
GraphRAG designs differ in how structures are constructed and how structural information is used during retrieval. 
In this work, we group GraphRAG approaches into four representative classes and select representative implementations for each class: 

\noindent \textbf{KG-based GraphRAG.}
In KG-based GraphRAG~\cite{Liu_LlamaIndex_2022}, a knowledge graph is constructed from text. 
Given a query, relevant entities are identified and aligned to nodes in the KG, and retrieval traverses multi-hop neighborhoods to collect relational triplets \textit{(head, relation, tail)} as evidence.
We consider two variants: 
{\bf (1)} {\it KG-GraphRAG (Triplets)}, which retrieves only triplets, and 
{\bf (2)} {\it KG-GraphRAG (Triplets+Text)}, which retrieves both triplets and their associated source text. 
We implement KG-GraphRAG using LlamaIndex~\cite{Liu_LlamaIndex_2022}.\footnote{{https://www.llamaindex.ai/}}

\noindent \textbf{Community-based GraphRAG.}
Community-based GraphRAG~\cite{edge2024local} further organizes the constructed KG into hierarchical communities using graph clustering algorithms. 
Each community is associated with a textual summary/report, where lower-level communities capture fine-grained information and higher-level communities provide increasingly abstract representations.
We evaluate two retrieval modes: 
{\bf Local Search} retrieves entity neighborhoods and lower-level community reports via entity matching, denoted as {\it Community-GraphRAG (Local)}; 
{\bf Global Search} retrieves high-level community summaries by semantic similarity, denoted as {\it Community-GraphRAG (Global)}.
We adopt the implementation of \citet{edge2024local}.\footnote{\url{https://microsoft.github.io/graphrag}}

\noindent \textbf{Text-centric graph-guided RAG.}
Text-centric graph-guided methods retain original text chunks as the primary retrieval units while leveraging graph structures to guide scoring or traversal. 
We select HippoRAG2~\cite{gutierrez2025rag} as a representative method. 
HippoRAG2 builds an entity-linked graph over chunks and retrieves query-relevant entities first, followed by the connected text chunks. 
Here, the graph acts as an auxiliary structure guiding chunk retrieval rather than the primary retrieval target.

\noindent \textbf{Hierarchical summary-based GraphRAG.}
Hierarchical summary-based methods construct multi-level hierarchical structures over text, where higher-level nodes represent progressively more abstract summaries of lower-level content. 
We adopt RAPTOR~\cite{sarthi2024raptor} as a representative method. 
RAPTOR recursively clusters text chunks and generates summaries at each level, enabling coarse-to-fine, multi-granular retrieval without relying on explicit KGs.

\subsection{Tasks}
We evaluate all methods on two representative text-based tasks: Question Answering and Query-based Summarization, covering both single-hop and multi-hop QA, as well as single-document and multi-document summarization. For single-document tasks, retrieval is restricted to the corresponding document, while for multi-document tasks, retrieval is performed over an index constructed from all documents.

\subsection{Unified Experimental Settings}
To ensure fair comparison across RAG and GraphRAG methods, we standardize core settings whenever applicable.

\noindent \textbf{Graph construction.}
For GraphRAG methods that require graph construction from text (e.g., KG-GraphRAG, Community-GraphRAG, and HippoRAG2), graphs are constructed using GPT-4o-mini; results with GPT-4o are reported in Appendix~\ref{app:graphconstruction}.

\noindent \textbf{Chunking.}
We segment documents into chunks of approximately \textit{256} tokens for all methods.

\noindent \textbf{Embedding model and retrieval budget.}
We embed queries, chunks, and graph information into a shared vector space using OpenAI's \texttt{text-embedding-ada-002} model~\cite{nussbaum2024nomic}, and retrieve top-$k$ candidates by semantic similarity, where $k{=}10$ by default. 

\noindent \textbf{Reranking.}
When reranking is enabled, we first retrieve the top $2k$ candidates and apply a cross-encoder reranker to reorder retrieved candidates and select the final top-$k$ evidence units. 
We use \texttt{BAAI/bge-reranker-large}~\cite{bge_embedding} as the reranker for all methods that support reranking, ensuring consistent reranking behavior across systems.

\noindent \textbf{Iterative retrieval.}
When iterative retrieval is enabled, we adopt IRCoT~\cite{trivedi2022interleaving}, which interleaves retrieval with intermediate reasoning steps.

\noindent \textbf{Generation backbones.}
To control for generation capacity, we use two open-source instruction-tuned LLMs of different sizes as generators: Llama-3.1-8B-Instruct and Llama-3.1-70B-Instruct~\cite{dubey2024llama}.

\section{Question Answering}
\label{sec:qa}
Question Answering (QA) is one of the most widely used tasks for evaluating the performance of RAG systems. To systematically assess the effectiveness of RAG and GraphRAG on general QA settings, we evaluate representative methods on widely used datasets and follow standard metrics used in prior work.

\subsection{Datasets and Evaluation Metrics}
To comprehensively evaluate the performance of GraphRAG on general QA tasks, we select four widely used datasets that cover different perspectives. For the single-hop QA task, we select the Natural Questions (NQ) dataset~\cite{kwiatkowski2019natural}. 
For the multi-hop QA task, we select HotPotQA~\cite{yang2018hotpotqa} and MultiHop-RAG~\cite{tang2024multihop} datasets. 
The MultiHop-RAG dataset categorizes queries into four types: Inference, Comparison, Temporal, and Null queries. To further analyze the performance of RAG and GraphRAG at a finer granularity, we also include NovelQA~\cite{wang2024novelqa}, which contains 21 different types of queries.
We use Precision (P), Recall (R), and F1-score as evaluation metrics for the NQ and HotPotQA datasets, while accuracy is used for the MultiHop-RAG and NovelQA datasets following their original papers.
\begin{table}[!htb]
\caption{Performance comparison (\%) on NQ and Hotpot.}
\label{tab:nq_llama8b}
\centering
\resizebox{\linewidth}{!}{
\begin{tabular}{l|ccc|ccc}
\toprule
\multirow{2}{*}{\textbf{Method}} 
& \multicolumn{3}{c|}{\textbf{NQ}} 
& \multicolumn{3}{c}{\textbf{Hotpot}} \\
\cmidrule{2-7}
& P & R & F1 & P & R & F1 \\
\midrule
RAG                        
& \textbf{71.70} & \textbf{63.93} & \textbf{64.78} 
& 62.32 & 60.47 & 60.04 \\

RaptorRAG
&66.06	&59.56	&60.04
&63.81	&61.46	&61.31 \\

KG-GraphRAG (Triplets only) 
& 40.09 & 33.56 & 34.28 
& 26.88 & 24.81 & 25.02 \\

KG-GraphRAG (Triplets+Text) 
& 58.36 & 48.93 & 50.27 
& 45.22 & 42.85 & 42.60 \\

Community-GraphRAG (Local)  
& \underline{69.48} & \underline{62.54} & \underline{63.01} 
& \underline{64.14} & \underline{62.08} & \underline{61.66} \\

Community-GraphRAG (Global) 
& 60.76 & 54.99 & 54.48 
& 45.72 & 47.60 & 45.16 \\

HippoRAG2 
& 67.25 & 60.42 & 61.03 
& \textbf{65.31} & \textbf{63.26} & \textbf{63.01} \\
\bottomrule
\end{tabular}
}
\end{table}

\begin{table}[htb]
\caption{Performance comparison (\%) on the MultiHop-RAG.}
\label{tab:multihop_llama8b}
\centering
\resizebox{\linewidth}{!}{
\begin{tabular}{l|ccccc}
\toprule
\textbf{Method} 
& \textbf{Inference} 
& \textbf{Comparison} 
& \textbf{Null} 
& \textbf{Temporal} 
& \textbf{Overall} \\
\midrule
RAG                        
& \textbf{92.16} & 57.59 & 96.01 & 30.70 & 67.02 \\

RaptorRAG
&\underline{91.91}	&55.26	&{90.03}	&45.28	&68.78 \\

KG-GraphRAG (Triplets only) 
& 55.76 & 22.55 & \textbf{98.67} & 18.70 & 41.24 \\

KG-GraphRAG (Triplets+Text) 
& 67.40 & 34.70 & \underline{97.34} & 17.15 & 48.51 \\

Community-GraphRAG (Local)  
& 86.89 & \underline{60.63} & 80.07 & \underline{50.60} & \underline{69.01} \\

Community-GraphRAG (Global) 
& 89.34 & \textbf{64.02} & 19.27 & \textbf{53.34} & 64.40 \\

HippoRAG2 
& 91.54 & 58.41 &85.71 & 49.91 & \textbf{70.27} \\
\bottomrule
\end{tabular}
}
\end{table}

\begin{table*}[!htb]
\caption{Performance comparison (\%) on the NovelQA dataset.}
\label{tab:novelqa}
\resizebox{\linewidth}{!}{
\begin{tabular}{ccccccccc|ccccccccc}
\toprule
\multicolumn{1}{l}{} 
& \multicolumn{8}{c|}{\textbf{RAG}} 
& \multicolumn{9}{c}{\textbf{RaptorRAG}} \\ 
\midrule
\multicolumn{1}{l}{} 
& chara & mean & plot & relat & settg & span & times & avg
& \multicolumn{1}{l}{} & chara & mean & plot & relat & settg & span & times & avg \\

mh & 68.75 & 52.94 & 58.33 & 75.28 & 92.31 & 64.00 & 33.96 & 47.34
   & mh & 60.42 & 70.59 & 63.89 & 65.17 & 92.31 & 52 & 38.24 & 48.17 \\
sh & 69.08 & 62.86 & 66.11 & 75.00 & 78.35 & - & - & 68.73
   & sh & 66.45 & 58.57 & 65.27 & 62.50 & 74.23 & - & - & 66.25 \\
dtl & 64.29 & 45.51 & 78.57 & 10.71 & 83.78 & - & - & 55.28
    & dtl & 62.86 & 48.88 & 80.36 & 28.57 & 78.38 & - & - & 57.72 \\
avg & 67.78 & 50.57 & 67.37 & 60.80 & 80.95 & 64.00 & 33.96 & 57.12
    & avg & 64.44 & 52.83 & 67.67 & 56.80 & 76.87 & 52 & 38.24 & 57.12 \\

\midrule
\multicolumn{1}{l}{} 
& \multicolumn{8}{c|}{\textbf{KG-GraphRAG (Triplets+Text)}} 
& \multicolumn{9}{c}{\textbf{Community-GraphRAG (Local)}} \\ 
\midrule
\multicolumn{1}{l}{} 
& chara & mean & plot & relat & settg & span & times & avg
& \multicolumn{1}{l}{} & chara & mean & plot & relat & settg & span & times & avg \\

mh & 52.08 & 52.94 & 44.44 & 55.06 & 69.23 & 64.00 & 28.61 & 38.37
   & mh & 68.75 & 64.71 & 55.56 & 67.42 & 92.31 & 52.00 & 35.83 & 47.01 \\
sh & 36.84 & 45.71 & 40.17 & 87.50 & 36.08 & - & - & 39.93
   & sh & 59.87 & 58.57 & 65.69 & 87.50 & 64.95 & - & - & 63.43 \\
dtl & 38.57 & 30.90 & 42.86 & 21.43 & 32.43 & - & - & 33.60
    & dtl & 54.29 & 37.64 & 62.50 & 25.00 & 70.27 & - & - & 46.88 \\
avg & 40.00 & 36.23 & 41.09 & 49.60 & 38.10 & 64.00 & 28.61 & 37.80
    & avg & 60.00 & 44.91 & 64.05 & 59.20 & 68.71 & 52.00 & 35.83 & 53.03 \\

\midrule
\multicolumn{1}{l}{} 
& \multicolumn{8}{c|}{\textbf{Community-GraphRAG (Global)}} 
& \multicolumn{9}{c}{\textbf{HippoRAG2}} \\ 
\midrule
\multicolumn{1}{l}{} 
& chara & mean & plot & relat & settg & span & times & avg
& \multicolumn{1}{l}{} & chara & mean & plot & relat & settg & span & times & avg \\

mh & 54.17 & 58.82 & 55.56 & 56.18 & 53.85 & 68 & 20.59 & 34.39
   & mh & 58.33 & 64.71 & 66.67 & 69.66 & 92.31 & 48 & 37.17 & 47.84 \\
sh & 45.39 & 50.00 & 55.65 & 87.50 & 38.14 & - & - & 49.65
   & sh & 65.79 & 65.71 & 64.44 & 62.50 & 72.16 & - & - & 66.25 \\
dtl & 28.57 & 29.78 & 32.14 & 87.50 & 40.54 & - & - & 30.89
    & dtl & 60.00 & 48.88 & 69.64 & 28.57 & 81.08 & - & - & 55.83 \\
avg & 42.59 & 36.98 & 51.66 & 52.00 & 40.14 & 68 & 20.59 & 39.17
    & avg & 62.96 & 54.34 & 65.56 & 60.00 & 76.19 & 48 & 37.17 & 56.54 \\

\bottomrule
\end{tabular}
}
\end{table*}


\subsection{QA Main Results}
\label{sec:qa_result}
We first compare the vanilla versions of RAG and GraphRAG variants. 
Unless otherwise specified, we report {main-paper results using Llama-3.1-8B-Instruct. 
We report results on NQ and HotpotQA, MultiHop-RAG, and NovelQA in Tables~\ref{tab:nq_llama8b}, \ref{tab:multihop_llama8b}, and \ref{tab:novelqa}, respectively.
We summarize key observations below:
\begin{enumerate}[leftmargin=*, itemsep=0pt, parsep=0pt]
    \item \textbf{RAG excels on detailed single-hop queries.} 
    RAG achieves strong performance on the single-hop benchmark NQ and on the single-hop (sh) and detail-oriented (dtl) subsets of NovelQA (Tables~\ref{tab:nq_llama8b} and~\ref{tab:novelqa}).
    \item \textbf{GraphRAG methods (e.g., HippoRAG2 and Community-GraphRAG (Local)) excel on multi-hop queries.} 
    They perform best on multi-hop QA benchmarks (HotPotQA and MultiHop-RAG) and remain competitive on the multi-hop (mh) subset of NovelQA (Tables~\ref{tab:nq_llama8b},~\ref{tab:multihop_llama8b}, and~\ref{tab:novelqa}).
    \item \textbf{Community-GraphRAG (Global) often struggles on QA.} 
    Global search retrieves high-level community summaries, which can lose fine-grained evidence and hurt detail-centric QA, as reflected on detail-oriented subsets in NovelQA. It also performs poorly on Null queries in MultiHop-RAG (ideally answered as \texttt{insufficient information}), suggesting increased hallucination risk. However, summary-level retrieval can be beneficial for Comparison and Temporal queries in MultiHop-RAG which require global information.
    
    \item \textbf{KG-based GraphRAG generally underperforms on QA due to limited graph coverage.} 
    KG-GraphRAG retrieves evidence primarily from extracted entities and relations, which can be incomplete and omit answer-critical information. In Appendix~\ref{app:sec:retrieve}, we show that only about 65.8\% of answer entities appear in the constructed KG for HotPotQA and 65.5\% for NQ, highlighting the sensitivity of KG-based retrieval to graph construction quality.
    
\end{enumerate}

\begin{figure}[!htb]
    \centering
    \includegraphics[width=\linewidth]{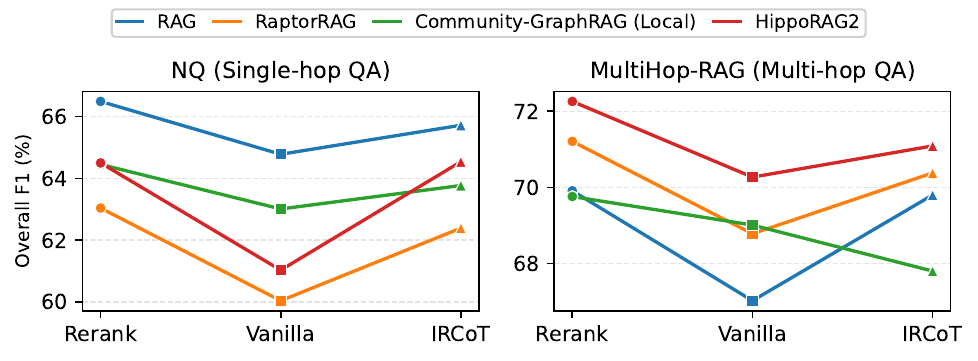}
    \caption{
        Overall QA performance (F1) under different inference strategies on
        NQ and MultiHop-RAG.
    }
    \label{fig:rerank_and_ircot}
\end{figure}

\subsection{QA with Reranking and Iterative Retrieval}
In addition to vanilla retrieval, we examine the impact of \emph{reranking} and \emph{iterative retrieval} on QA performance. We conduct experiments on \textbf{NQ} and \textbf{MultiHop-RAG}, representing single-hop and multi-hop QA scenarios, respectively. Figure~\ref{fig:rerank_and_ircot} summarizes overall performance under different inference strategies.

Across both datasets, reranking and iterative retrieval generally improve performance for all methods compared to vanilla inference, indicating that inference-time enhancements provide gains beyond the underlying retrieval architecture. On \textbf{NQ}, both reranking and iterative retrieval (IRCoT) yield consistent improvements, suggesting that such refinements can be beneficial even for predominantly single-hop QA. Importantly, our earlier conclusion remains unchanged: RAG still performs better on single-hop, detail-oriented questions, even when equipped with reranking or iterative retrieval. On the \textbf{MultiHop-RAG} benchmark, the gains from inference-time strategies are more pronounced. Both reranking and IRCoT lead to larger absolute improvements than on NQ, highlighting the value of progressive evidence refinement in multi-hop settings. Under these enhanced inference strategies, GraphRAG methods also typically outperform RAG. One exception is Community-GraphRAG (Local) with IRCoT, which exhibits notably low performance on \texttt{NULL} queries, despite improvements on other categories. This observation remains consistent with our main findings.
Overall, reranking and iterative retrieval are complementary to both RAG and GraphRAG, and are particularly important for multi-hop QA. 

\begin{figure}[htb]
    \centering
    \begin{subfigure}[b]{0.35\columnwidth}
        \centering
        \includegraphics[width=\textwidth]{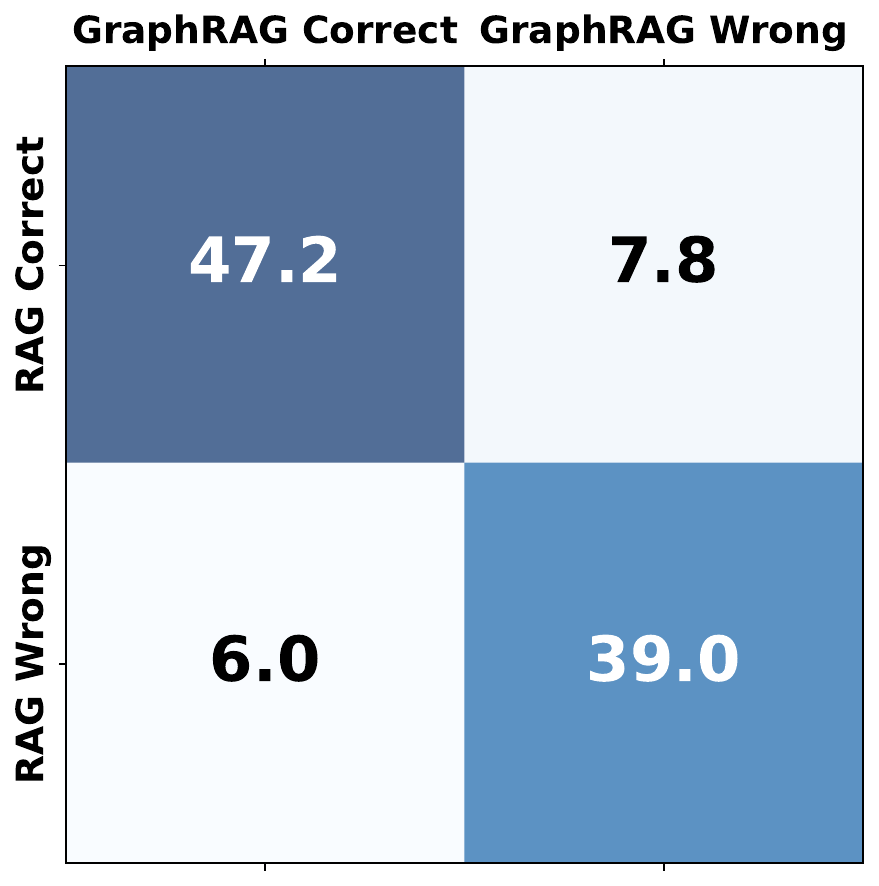}
        \caption{NQ}
    \end{subfigure}
   \hspace{0.01\textwidth}
    \begin{subfigure}[b]{0.35\columnwidth}
        \centering
        \includegraphics[width=\textwidth]{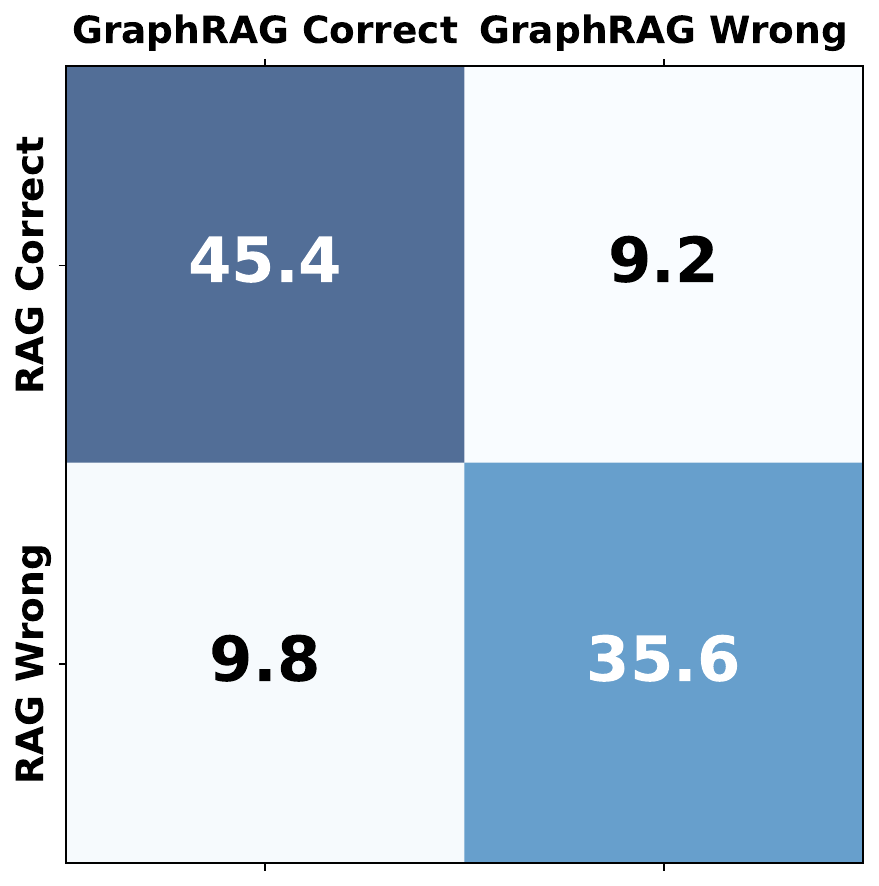}
        \caption{HotpotQA}
    \end{subfigure}

    \vspace{0.1cm}

    \begin{subfigure}[b]{0.35\columnwidth}
        \centering
        \includegraphics[width=\textwidth]{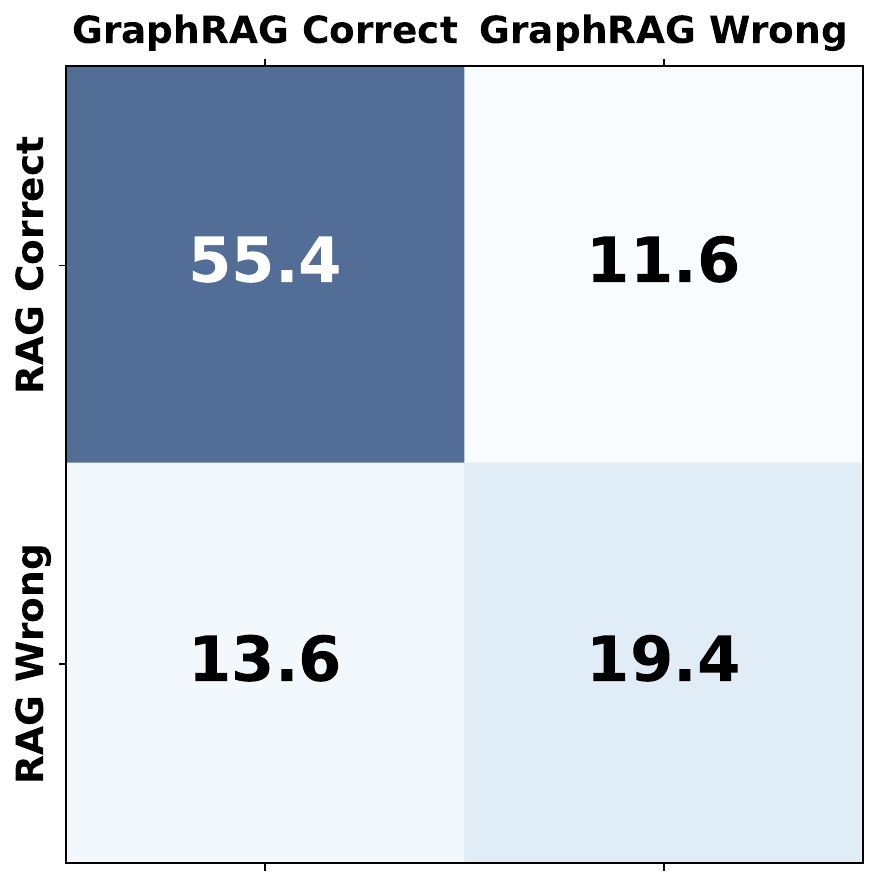}
        \caption{MultiHop-RAG}
    \end{subfigure}
    \hspace{0.01\textwidth}
    \begin{subfigure}[b]{0.35\columnwidth}
        \centering
        \includegraphics[width=\textwidth]{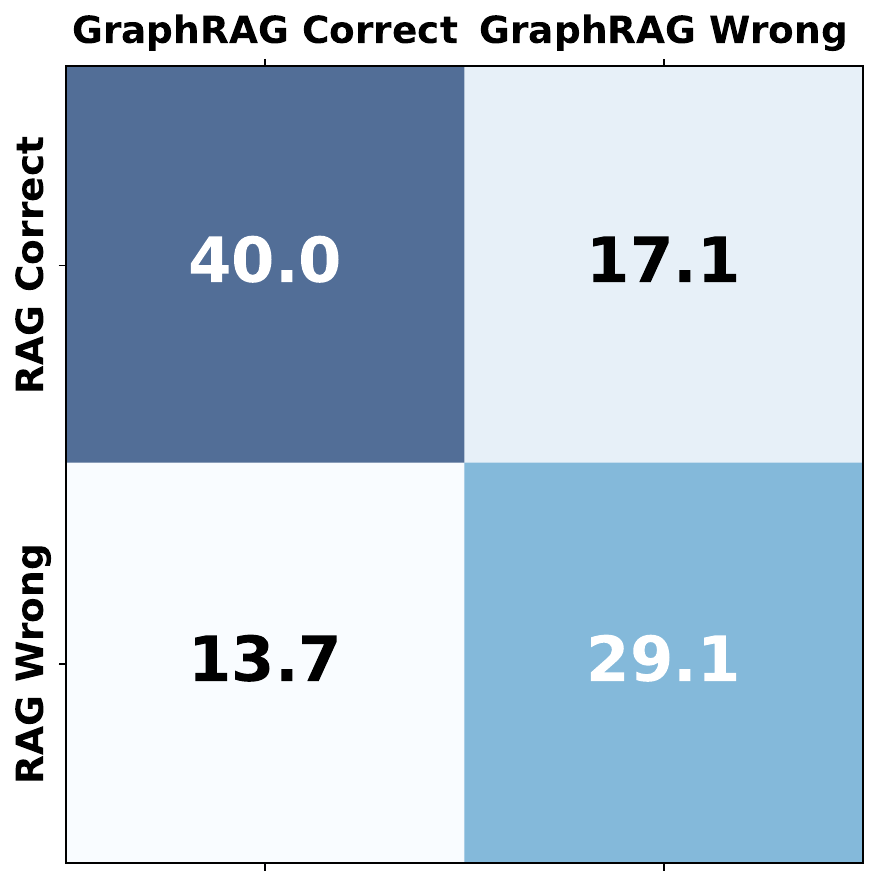}
        \caption{NovelQA}
    \end{subfigure}
    \caption{Confusion matrices comparing GraphRAG and RAG correctness across datasets using Llama~3.1-8B.}
    \label{fig:confusion}
        \vspace{-0.2in}
\end{figure}

\subsection{Comparative QA Analysis}
In this section, we provide a detailed comparison of RAG and GraphRAG, with a focus on their strengths and weaknesses. We consider vanilla RAG and Community-GraphRAG (Local), and refer to the latter simply as GraphRAG, as it exhibits performance comparable to RAG in our experiments. 
We partition all queries into four categories: {\bf (1)} queries correctly answered by both methods, {\bf (2)} queries correctly answered only by RAG (RAG-only), {\bf (3)} queries correctly answered only by GraphRAG (GraphRAG-only), and {\bf (4)} queries incorrectly answered by both methods.

The confusion matrices representing these four groups using the Llama 3.1-8B model are shown in Figure~\ref{fig:confusion}. Notably, the proportions of queries correctly answered exclusively by GraphRAG and RAG are significant. For example, 13.6\% of queries are GraphRAG-only, while 11.6\% are RAG-only on MultiHop-RAG dataset. This phenomenon highlights the complementary properties of RAG and GraphRAG.
Therefore, {\it leveraging their unique advantages has the potential to improve overall performance}.



\subsection{Improving QA Performance}
\label{sec:improve_qa}
Building on the complementary properties of RAG and GraphRAG, we investigate the following two strategies to enhance overall QA performance.

\paragraph{Strategy 1: RAG vs. GraphRAG Selection}
In Section~\ref{sec:qa_result}, we observe that RAG generally performs well on single-hop queries and those requiring detailed information, while GraphRAG (Community-GraphRAG (Local)) excels in multi-hop queries that require reasoning. Therefore, we hypothesize that RAG is well-suited for fact-based queries, which rely on direct retrieval and detailed information, whereas GraphRAG is more effective for reasoning-based queries that involve chaining multiple facts together. Therefore, given a query, we employ a classification mechanism to determine whether it is fact-based or reasoning-based. Each query is then assigned to either RAG or GraphRAG based on the classification results. Specifically, we leverage the in-context learning ability of LLMs for classification~\cite{dong2022survey, wei2023larger}. In this strategy, either RAG or GraphRAG is selected for each query, and we refer to this strategy as \textbf{Selection}.

\paragraph{Strategy 2: RAG and GraphRAG Integration}
We further explore an \textbf{Integration} strategy that jointly leverages the complementary retrieval behaviors of RAG and GraphRAG. For each query, both methods retrieve relevant information in parallel, and the retrieved contexts are concatenated and fed into the generator to produce the final answer.
We evaluate the effectiveness of the two proposed strategies on all selected datasets. For MultiHop-RAG and NovelQA, we report overall accuracy, while for NQ and HotPotQA, we use F1 score as the evaluation metric. The results are summarized in Figure~\ref{fig:qa_improve}.

\begin{figure*}[!htb]
    \centering
    \begin{subfigure}[b]{0.4\textwidth}
        \centering
        \includegraphics[width=\textwidth]{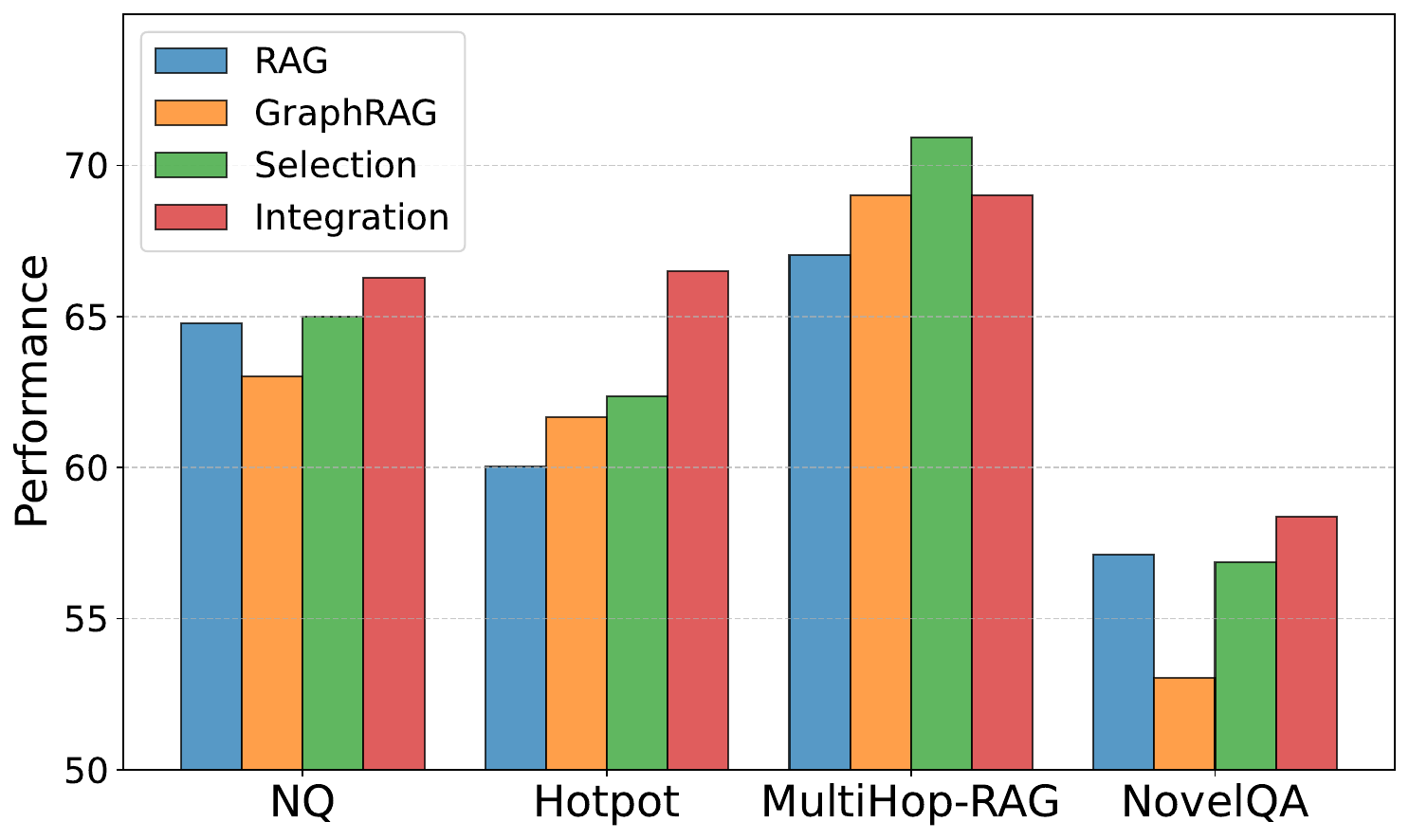}
        \caption{Llama3.1-8B}
        \label{fig:qa_improve_8b}
    \end{subfigure}
    \hspace{0.01\textwidth}
    \begin{subfigure}[b]{0.4\textwidth}
        \centering
        \includegraphics[width=\textwidth]{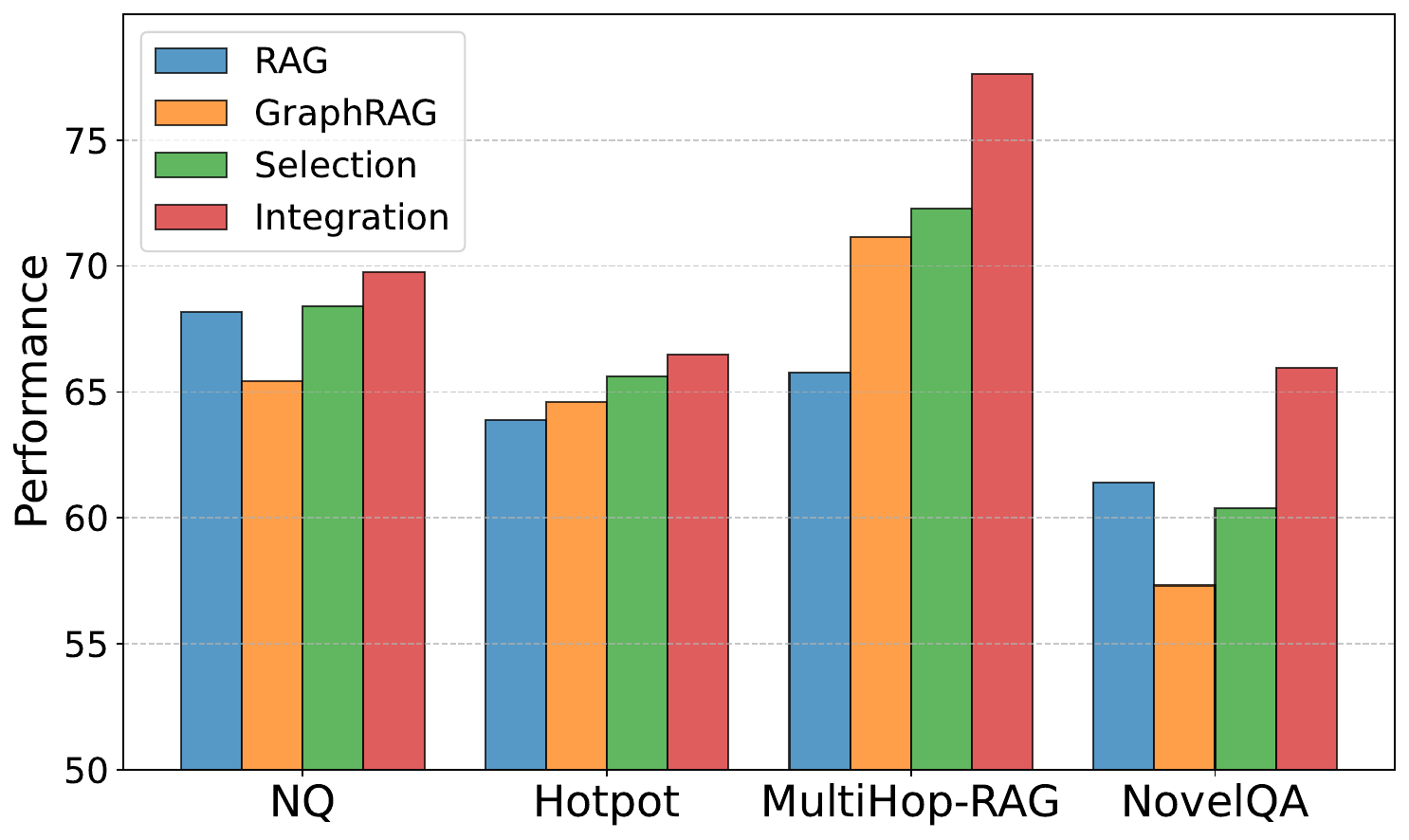}
        \caption{Llama3.1-70B}
        \label{fig:qa_improve_70b}
    \end{subfigure}
    \caption{Overall QA performance comparison of different methods.}
    \label{fig:qa_improve}
\end{figure*}

Overall, both strategies consistently improve QA performance across datasets. For instance, on MultiHop-RAG with Llama~3.1-70B, the Selection and Integration strategies improve the best baseline by 1.1\% and 6.4\%, respectively. When comparing the two strategies, Integration generally achieves higher performance than Selection.
However, Selection processes each query using only a single method, making it more computationally efficient. In contrast, Integration requires running both RAG and GraphRAG for every query, leading to higher computational cost.

\subsection{Computation and Storage Analysis}
In this subsection, we analyze the computational and storage trade-offs among RAG, KG-GraphRAG, and Community-GraphRAG. Specifically, we report construction time, retrieval latency, and storage footprint on the MultiHop-RAG dataset. The results are summarized in Table~\ref{tab:time:multi}.
From the results, we have the following observations:
\begin{itemize}[leftmargin=1em]
\item \textbf{Construction time:} Both GraphRAG variants incur substantially higher construction cost than RAG, as they require additional graph construction and preprocessing.
\item \textbf{Retrieval time:} KG-GraphRAG exhibits the highest retrieval latency, primarily due to LLM-based entity expansion and multi-step graph traversal. In contrast, Community-GraphRAG achieves the lowest retrieval latency by relying on direct community-level matching, even outperforming vanilla RAG.
\item \textbf{Storage:} Community-GraphRAG requires the largest storage footprint, as it stores both community representations and associated summaries. KG-GraphRAG is more storage-efficient than Community-GraphRAG and RAG, reflecting a trade-off between information richness and storage cost.
\end{itemize}


\begin{table}[!htb]
\centering
\caption{The time and storage analysis on MultiHop-RAG.}
\label{tab:time:multi}
\resizebox{\linewidth}{!}{
\begin{tabular}{c|ccc}
\hline
Method             & Construction Time (s) & Retrieval time (s) & Storage \\ \hline
RAG                & 135               & 1724          & 127MB   \\
KG-GraphRAG        & 7702              & 14434         & 117MB   \\
Community-GraphRAG & 5560              & 1249          & 165MB   \\ \hline
\end{tabular}
}
\end{table}

We also report token-matched results across different methods in Appendix~\ref{app:sec:token_matched}.

\subsection{Graph Construction Model}
In this subsection, we analyze how graph construction quality affects GraphRAG performance. We focus on Community-GraphRAG (Local) as a representative GraphRAG variant and vary the LLM used for graph construction, while keeping the retrieval and generation components fixed. We evaluate downstream QA performance across multiple tasks and datasets, with detailed results reported in Appendix~\ref{app:graphconstruction}. Table~\ref{app:tab:cons2} summarizes results on MultiHop-RAG across different query categories.

\begin{table}[!htb] 
\centering 
\caption{Impact of graph construction models on the MultiHop-RAG dataset using Llama~3.1--70B-Instruct.} 
\label{app:tab:cons2} 
\resizebox{\linewidth}{!}{ 
\begin{tabular}{c|ccccc} \hline 
\textbf{Graph Construction} & Inference & Comparison & NULL & Temporal & Overall \\ \hline 
None (RAG) & 94.85 & 56.31 & 91.36 & 25.73 & 65.77 \\ 
GPT-4o-mini & 92.03 & 60.16 & 88.70 & 49.06 & 71.17 \\ 
GPT-4o & 93.63 & 66.59 & 81.06 & 58.49 & 75.08 \\ \hline 
\end{tabular} } 
\end{table}

We observe that stronger graph construction models consistently improve QA performance, especially on reasoning-intensive queries such as Comparison and Temporal. While omitting explicit graph construction (i.e., RAG) yields strong performance on Inference and NULL queries, it performs poorly on multi-hop reasoning tasks. In contrast, graphs constructed using more capable LLMs, such as GPT-4o, substantially improve performance on these challenging categories, leading to the highest overall accuracy.
These results indicate that GraphRAG performance is sensitive to graph construction quality. However, stronger construction models also incur higher computational cost, highlighting a trade-off between graph quality and system efficiency when selecting LLMs for graph construction.

\noindent\textbf{Summary.}
Across our QA experiments, we find that RAG and GraphRAG exhibit complementary strengths rather than a clear dominance. RAG consistently performs well on single-hop, fact-centric queries that require precise retrieval of detailed information, while GraphRAG excels on reasoning-intensive, multi-hop queries by explicitly modeling relationships among entities. However, these benefits come with different computational and storage trade-offs, and GraphRAG performance is further influenced by the quality of graph construction. Motivated by these observations, we explore selection and integration strategies that combine the strengths of both paradigms, leading to consistent improvements in overall QA performance. Together, these results suggest that effective QA systems should adaptively balance retrieval precision, reasoning capability, and system efficiency, rather than relying on a single retrieval paradigm.

\section{Query-Based Summarization}
\label{sec:summ}

\begin{table*}[htb]
\caption{The performance of query-based single document summarization task using Llama 3.1-8B.}
\label{tab:summ_single}
\vspace{-0.1in}
\resizebox{0.9\linewidth}{!}{
\begin{tabular}{l|cccccc|cccccc}
\toprule
\multirow{4}{*}{\textbf{Method}}      
& \multicolumn{6}{c|}{\textbf{SQuALITY}}                                                                           
& \multicolumn{6}{c}{\textbf{QMSum}} \\
\cmidrule{2-13}
& \multicolumn{3}{c}{\textbf{ROUGE-2}}                     
& \multicolumn{3}{c|}{\textbf{BERTScore}}                        
& \multicolumn{3}{c}{\textbf{ROUGE-2}}                    
& \multicolumn{3}{c}{\textbf{BERTScore}} \\ 
\cmidrule{2-13}
& P & R & F1 & P & R & F1 & P & R & F1 & P & R & F1 \\
\midrule
RAG                        
& 15.09 & 8.74 & 10.08 
& 74.54 & 81.00 & 77.62 
& \underline{21.50} & {3.80} & 6.32 
& {81.03} & 84.45 & {82.69} \\

RaptorRAG                  
& 14.88 & 8.42 & 9.81 
& 74.55 & 81.20 & 77.71 
& 20.38 & \textbf{4.17} & \textbf{6.68} 
& \textbf{81.64} & \underline{84.57} & \textbf{83.07} \\

KG-GraphRAG (Triplets only) 
& 11.99 & 6.16 & 7.41 
& 82.46 & 84.30 & 83.17 
& 13.71 & 2.55 & 4.15 
& 80.16 & 82.96 & 81.52 \\

KG-GraphRAG (Triplets+Text) 
& 15.00 & \textbf{9.48} & \underline{10.52} 
& \textbf{84.37} & \textbf{85.88} & \textbf{84.92} 
& 16.83 & 3.32 & 5.38 
& 80.92 & 83.64 & 82.25 \\

Community-GraphRAG (Local)  
& \textbf{15.82} & 8.64 & 10.10 
& \underline{83.93} & \underline{85.84} & \underline{84.66} 
& 20.54 & 3.35 & 5.64 
& 80.63 & 84.13 & 82.34 \\

Community-GraphRAG (Global) 
& 10.23 & 6.21 & 6.99 
& 82.68 & 84.26 & 83.30 
& 10.54 & 1.97 & 3.23 
& 79.79 & 82.47 & 81.10 \\

HippoRAG2                   
& 15.07 & 8.95 & 10.20 
& 74.60 & 81.24 & 77.75 
& 21.35 & \underline{4.01} & \underline{6.60} 
& \underline{81.44} & \textbf{84.63} & \underline{83.00} \\

Integration                
& \underline{15.69} & \underline{9.32} & \textbf{10.67} 
& 74.56 & 81.22 & 77.73 
& \textbf{21.97} &{3.80} & {6.34} 
& 80.89 & {84.47} & 82.63 \\ 
\bottomrule
\end{tabular}
}
\end{table*}

\begin{table*}[htb]
\caption{The performance of query-based multiple document summarization task using Llama3.1-8B.}
\label{tab:summ_multi}
\vspace{-0.1in}
\resizebox{0.9\linewidth}{!}{
\begin{tabular}{l|cccccc|cccccc}
\toprule
\multirow{4}{*}{\textbf{Method}}                   
& \multicolumn{6}{c|}{\textbf{ODSum-story}}                                                                        
& \multicolumn{6}{c}{\textbf{ODSum-meeting}}                                                                       \\
\cmidrule{2-13}
& \multicolumn{3}{c}{\textbf{ROUGE-2}}                    
& \multicolumn{3}{c|}{\textbf{BERTScore}}                        
& \multicolumn{3}{c}{\textbf{ROUGE-2}}                    
& \multicolumn{3}{c}{\textbf{BERTScore}}                         \\
\cmidrule{2-13}
& P & R & F1 & P & R & F1 & P & R & F1 & P & R & F1 \\
\midrule
RAG                        
& \underline{15.39} & 8.44 & \underline{9.81} 
& \textbf{83.87} & \textbf{85.74} & \underline{84.57} 
& 15.50 & \textbf{6.43} & \textbf{8.77} 
& \textbf{83.12} & \textbf{85.84} & \textbf{84.45} \\

RaptorRAG                  
& 14.69 & \underline{8.47} & 9.62 
& \textbf{83.87} & 85.76 & \textbf{84.58} 
& 14.85 & \underline{6.21} & 8.44 
& 82.66 & 85.52 & 84.06 \\

KG-GraphRAG (Triplets only) 
& 11.02 & 5.56 & 6.62 
& 82.09 & 83.91 & 82.77 
& 11.64 & 4.87 & 6.58 
& 81.13 & 84.32 & 82.69 \\

KG-GraphRAG (Triplets+Text) 
& 9.19 & 5.82 & 6.22 
& 79.39 & 83.30 & 81.03 
& 11.97 & 4.97 & 6.72 
& 81.50 & 84.41 & 82.92 \\

Community-GraphRAG (Local)  
& 13.84 & 7.19 & 8.49 
& 83.19 & 85.07 & 83.90 
& 15.65 & 5.66 & 8.02 
& 82.44 & 85.54 & 83.96 \\

Community-GraphRAG (Global) 
& 9.40 & 4.47 & 5.46 
& 81.46 & 83.54 & 82.30 
& 11.44 & 3.89 & 5.59 
& 81.20 & 84.50 & 82.81 \\

HippoRAG2                   
& \textbf{15.56} & 8.43 & \textbf{9.82} 
& 83.70 & \underline{85.71} & {84.46} 
& \textbf{15.91} & 6.09 & \underline{8.51} 
& 82.43 & 85.55 & 83.95 \\

Integration                    
& 14.77 & \textbf{8.55} & 9.53 
& \underline{83.73} & 85.56 & 84.40 
& \underline{15.69} & {6.15} & \underline{8.51} 
& \underline{82.87} & \underline{85.81} & \underline{84.31} \\ 
\bottomrule
\end{tabular}
}
\end{table*}

Query-based summarization is a widely used benchmark for evaluating retrieval-augmented generation (RAG) systems~\cite{ram2023context, yu2023augmentation}. Recent work has also demonstrated the potential of GraphRAG for summarization tasks~\cite{edge2024local}. However, \citet{edge2024local} focus primarily on global summarization and rely on LLM-as-a-Judge~\cite{zheng2023judging} for evaluation. As shown in Section~\ref{sec:summ_analysis}, LLM-as-a-Judge introduces position bias in summarization evaluation, which can compromise result reliability. Despite the growing interest in GraphRAG, a systematic comparison between RAG and GraphRAG on general query-based summarization tasks across widely used datasets remains unexplored. To address this gap, we conduct a comprehensive evaluation in this section using standard benchmarks and evaluation metrics.

\subsection{Datasets and Evaluation Metrics}
We evaluate RAG and GraphRAG on four widely used query-based summarization datasets: two single-document datasets, SQuALITY~\cite{wang2022squality} and QMSum~\cite{zhong2021qmsum}, and two multi-document datasets, ODSum-story and ODSum-meeting~\cite{zhou2023odsum}.
Unlike the LLM-generated global queries used in the unreleased datasets of \citet{edge2024local}, most queries in the selected datasets focus on specific roles or events. Since these datasets contain one or more human-written ground truth summaries for each query, we leverage ROUGE-2~\cite{lin2004rouge} and BERTScore~\cite{zhang2019bertscore} as evaluation metrics to measure lexical and semantic similarity between the predicted and ground truth summaries.

\begin{figure*}[!htb]
    \centering
    \begin{subfigure}[b]{0.24\textwidth}
        \centering
        \includegraphics[width=\textwidth]{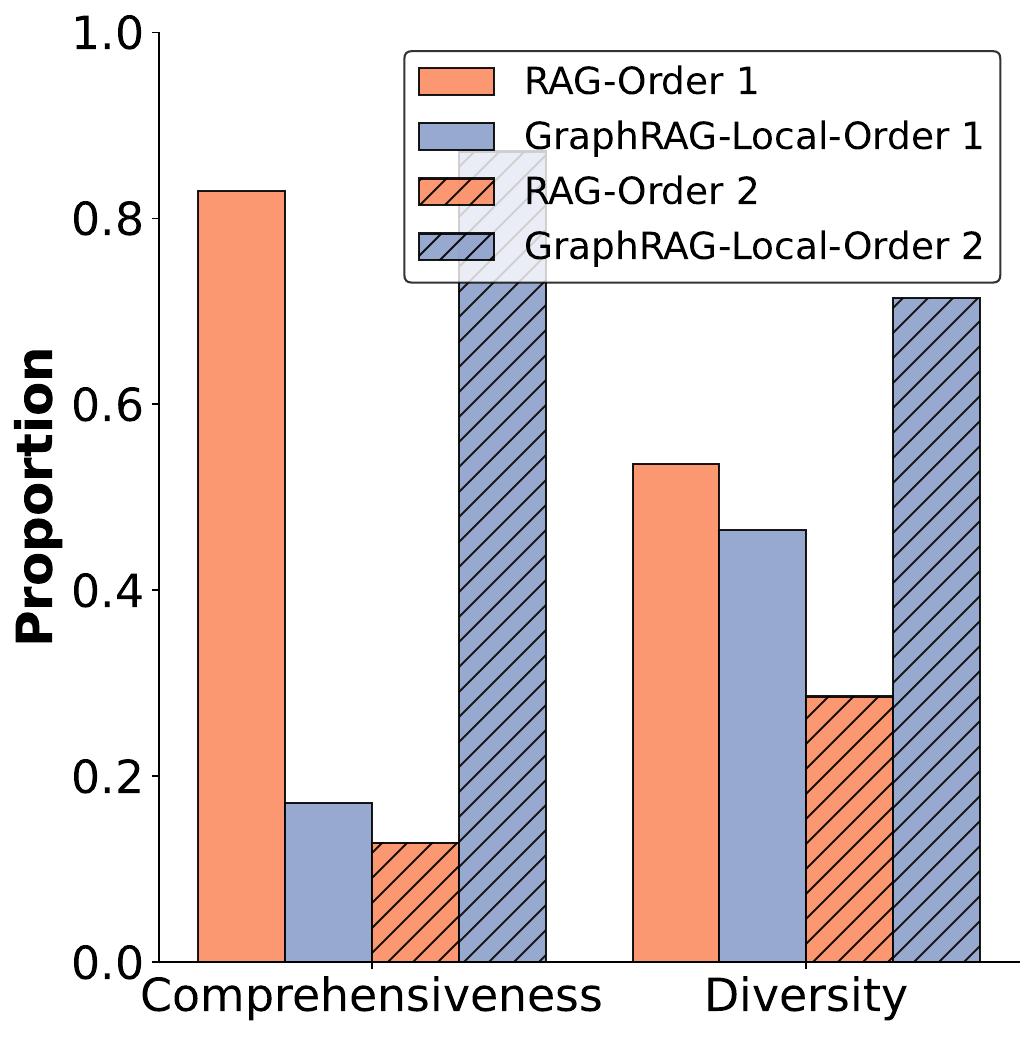}
        \caption{QMSum Local}
        \label{fig:qmsum_local}
    \end{subfigure}
    \hfill
    \begin{subfigure}[b]{0.24\textwidth}
        \centering
        \includegraphics[width=\textwidth]{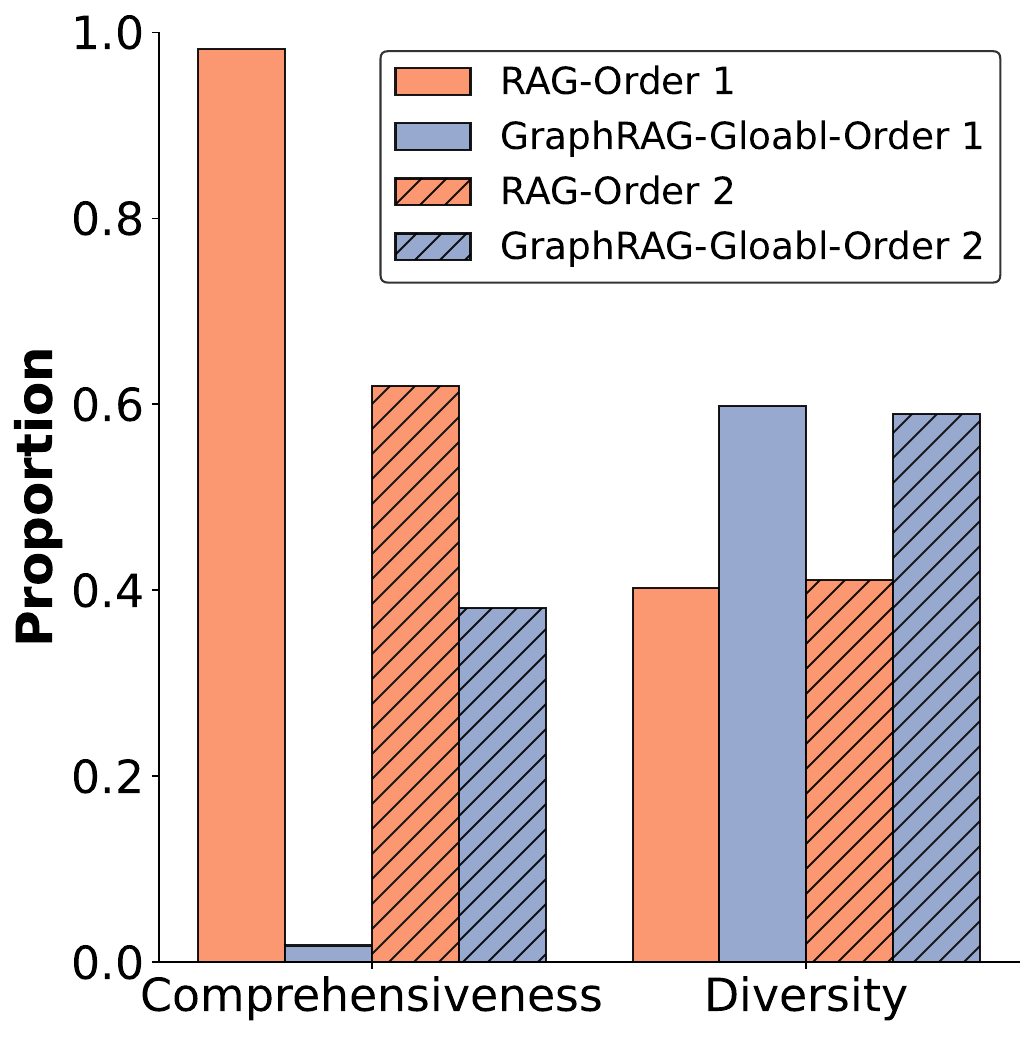}
        \caption{QMSum Global}
        \label{fig:qmsum_global}
    \end{subfigure}
    \hfill
    \begin{subfigure}[b]{0.24\textwidth}
        \centering
        \includegraphics[width=\textwidth]{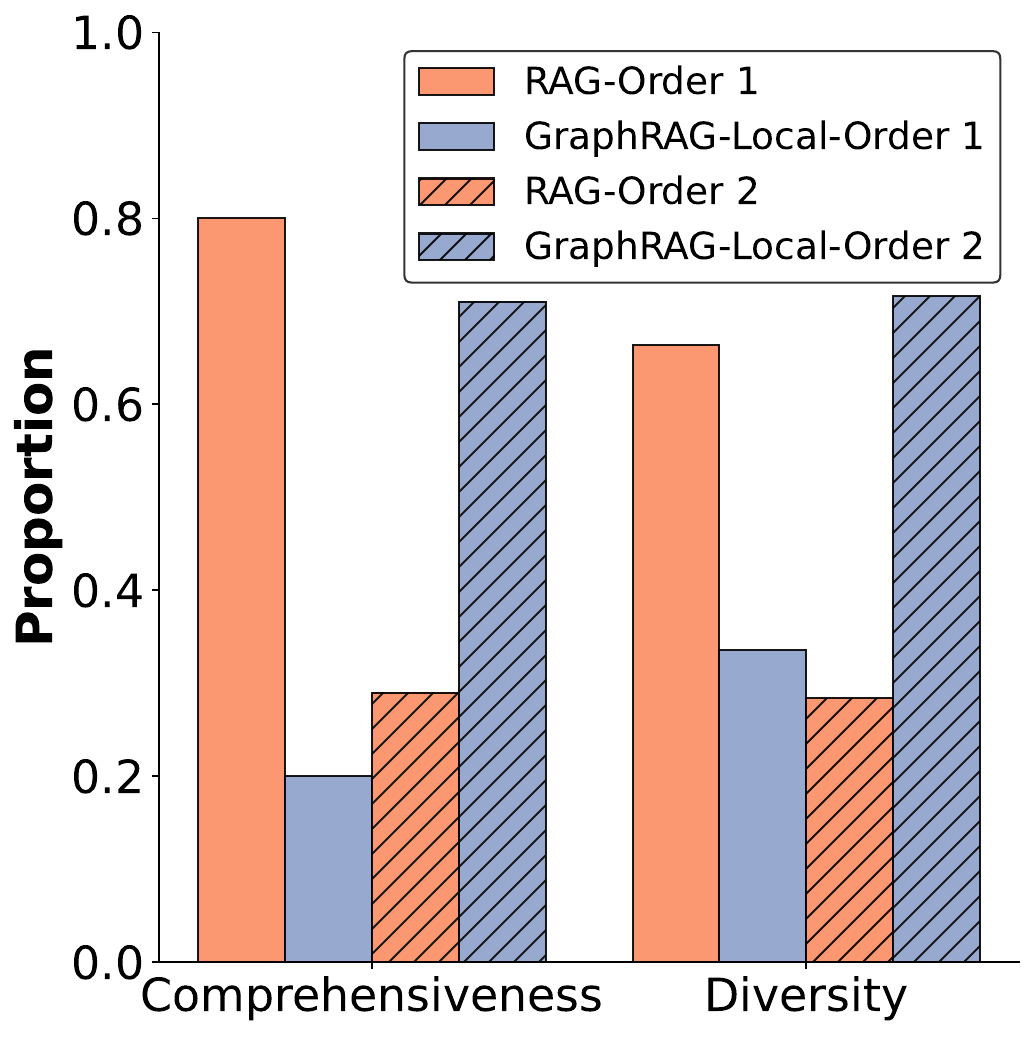}
        \caption{ODSum-story Local}
        \label{fig:odsum_local}
    \end{subfigure}
    \hfill
    \begin{subfigure}[b]{0.24\textwidth}
        \centering
        \includegraphics[width=\textwidth]{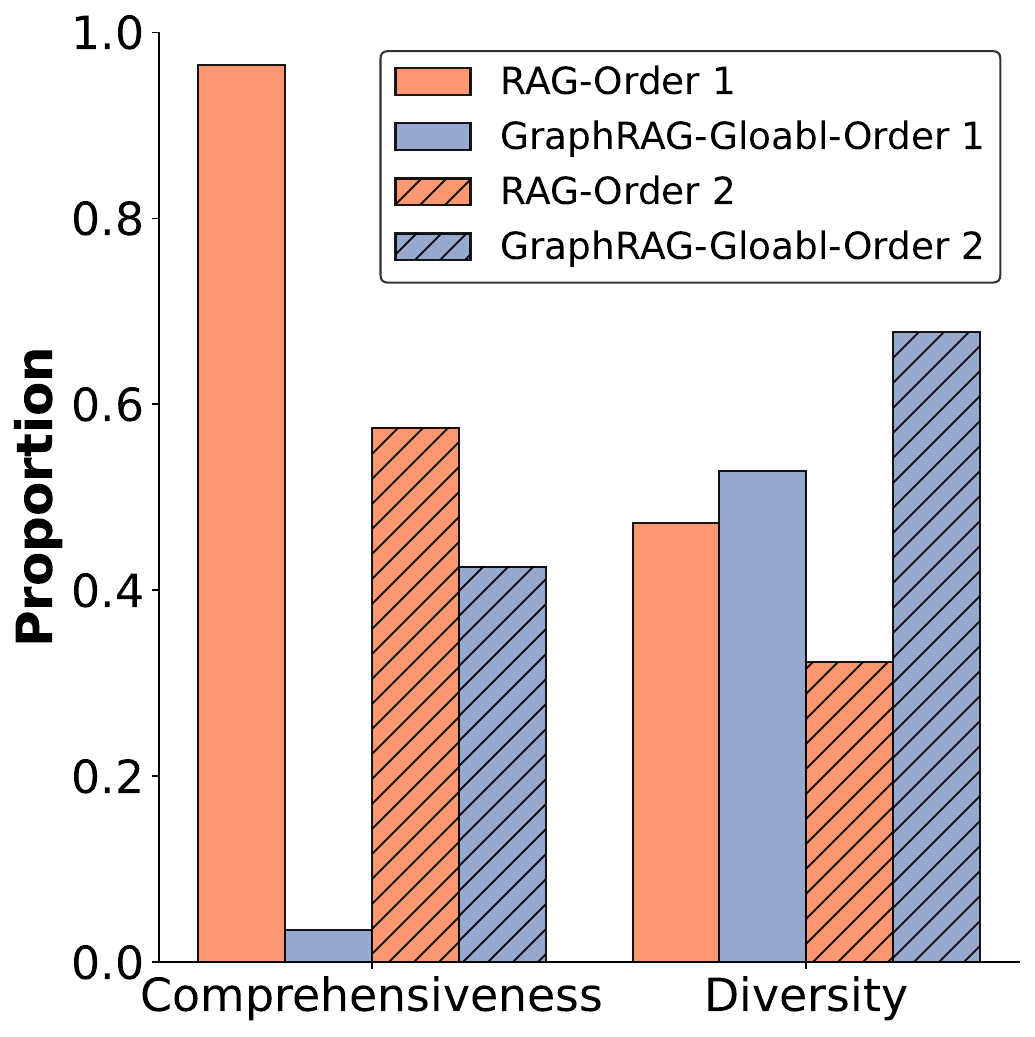}
        \caption{ODSum-story Global}
        \label{fig:odsum_global}
    \end{subfigure}
    \vspace{-0.1in}
    \caption{Comparison of LLM-as-a-Judge evaluations for RAG and GraphRAG. "Local" refers to the evaluation of RAG vs. GraphRAG-Local, while "Global" refers to RAG vs. GraphRAG-Global. }
    \label{fig:summ_comparision}
    \vspace{-0.15in}
\end{figure*}

\subsection{Summarization Experimental Results}
\label{sec:summ_result}

We evaluate on vanilla RAG and GraphRAG methods, along with the Integration strategy discussed in Section~\ref{sec:improve_qa}. 
The results of Llama3.1-8B model on Query-based single document summarization and multiple document summarization are shown in Table~\ref{tab:summ_single} and Table~\ref{tab:summ_multi}, respectively. 
Based on these results, we can make the following observations: 
\begin{enumerate}[leftmargin=*, itemsep=0pt, parsep=-1pt]
    \item {\bf RAG, RaptorRAG, and HippoRAG2 generally performs well on query-based summarization tasks}, primarily because they retrieve original text chunks that are more closely aligned with ground truth.
    \item {\bf KG-based GraphRAG benefit from combining triplets with their corresponding text}. This improves performance by incorporating more details, making predictions closer to the human-written ground truth summaries.
    \item {\bf Community-based GraphRAG performs better with the Local search method}. Local search retrieves entities, relations, and low-level communities, while the Global search method retrieves only high-level summaries. This demonstrates the importance of detailed information in the selected datasets.
    \item {\bf The Integration strategy often performs comparably to RAG alone}, suggesting that simply concatenating RAG and GraphRAG evidence does not reliably improve alignment with detailed ground-truth summaries.
\end{enumerate}

\subsection{Position Bias in Existing Evaluation}
\label{sec:summ_analysis}
Based on the results in Section~\ref{sec:summ_result}, Community-based GraphRAG, particularly with global search, generally underperforms RAG on the selected datasets. This observation contrasts with the findings of \citet{edge2024local}, where Community-based GraphRAG with global search outperformed both local search and RAG.

This discrepancy can be attributed to two key differences between our evaluation and that of \citet{edge2024local}. First, their study focuses on global summarization, which aims to capture high-level information from an entire corpus, whereas the datasets used in our evaluation involve queries targeting specific roles or events. Second, \citet{edge2024local} evaluate performance using LLM-as-a-Judge without ground-truth references, while we evaluate generated summaries against human-written references using ROUGE and BERTScore. These reference-based metrics emphasize factual coverage and fine-grained details, which may favor different retrieval behaviors.

To further investigate this difference, we follow \citet{edge2024local} and evaluate RAG and GraphRAG using LLM-as-a-Judge from two perspectives: \emph{Comprehensiveness} and \emph{Diversity}. Comprehensiveness measures how well a summary covers the details required by the query, while Diversity assesses whether the summary provides a broad and globally inclusive view. 
For each query, we present summaries generated by RAG and GraphRAG to the LLM and ask it to select the better one under each criterion.

To examine potential position effects in LLM-as-a-Judge evaluations, we consider two presentation orders: Order~1 (O1), where the RAG summary appears first, and Order~2 (O2), where the GraphRAG summary appears first. For each setting, we report the proportion of times each method is preferred by the LLM, where a higher proportion indicates stronger performance under the given presentation order.
Figure~\ref{fig:summ_comparision} presents results comparing RAG with GraphRAG (Local) and GraphRAG (Global) on the QMSum and ODSum-story datasets, using GPT-4o as the judge. We make two key observations. 
First, \textbf{position bias~\cite{shi2024judging, wang2024eliminating} is clearly present in LLM-as-a-Judge evaluations for summarization}, as reversing the order of presented summaries leads to substantially different, and in some cases opposite, judgments. This effect is especially pronounced for the comparison between RAG and GraphRAG (Local), as shown in Figures~\ref{fig:qmsum_local} and~\ref{fig:odsum_local}. 
Second, in comparisons between RAG and GraphRAG (Global), RAG is consistently preferred in terms of Comprehensiveness, while GraphRAG (Global) is favored for Diversity (Figures~\ref{fig:qmsum_global} and~\ref{fig:odsum_global}). This suggests that \textbf{Community-based GraphRAG with global search emphasizes corpus-level coverage, whereas RAG is more effective at capturing fine-grained, query-specific details}.

We further evaluate stronger LLM judge models, such as GPT-5.4, with results reported in the Appendix. The findings remain consistent. Position bias persists across judge models, although its severity varies: GPT-4o is highly sensitive to presentation order, while GPT-5.4 reduces but does not eliminate this bias. Moreover, the main conclusions remain stable: GraphRAG (Global) improves Diversity, while RAG achieves better Comprehensiveness due to its finer-grained evidence retrieval. We examine graph construction with different LLMs for summarization tasks in Appendix~\ref{app:graphconstruction}.

\noindent\textbf{Summary.}
Across query-based summarization benchmarks, RAG and GraphRAG exhibit different generation characteristic. Under reference-based metrics, RAG typically better matches detailed, query-specific ground-truth summaries, whereas GraphRAG, especially community-based global retrieval, tends to produce more corpus-level and diverse summaries that can deviate from fine-grained details. We also find that LLM-as-a-Judge evaluation can be sensitive to presentation order, raising reliability concerns for judge-based comparisons. Overall, these results highlight the need to balance detail fidelity, diversity, evaluation reliability, and system cost when applying (Graph)RAG to query-based summarization.

\section{Additional Evaluation and Token Cost Analysis}
\label{sec:additional_eval_cost}

To further examine whether our findings generalize beyond standard QA and summarization, we evaluate representative methods on two additional settings: LongBench-v2~\cite{bai2025longbench} for long-context reasoning and LoCoMo~\cite{maharana2024evaluating} for long-term memory question answering of LLM agent systems. These benchmarks require models to retrieve, aggregate, and reason over evidence from long documents or multi-session histories. We also provide an end-to-end token cost analysis to better understand the practical trade-offs of different retrieval strategies.

\begin{table}[t]
\centering
\caption{Performance of different methods on LongBench-v2 and LoCoMo.}
\label{tab:long_context_memory_results}
\resizebox{\linewidth}{!}{
\begin{tabular}{llccc}
\hline
Dataset & Category & RAG & RaptorRAG & HippoRAG2 \\
\hline
LongBench-v2 & Single-Document & 37.5 & \textbf{41.2} & 33.8 \\
LongBench-v2 & Multi-Document & 35.1 & 35.1 & \textbf{40.3} \\
LongBench-v2 & Long In-context Learning & \textbf{41.7} & 37.5 & 39.6 \\
\hline
LoCoMo & Single-session & 0.5955 & 0.5839 & \textbf{0.6149} \\
LoCoMo & Multi-session & 0.3362 & 0.3167 & \textbf{0.3607} \\
LoCoMo & Temporal & 0.1638 & 0.1617 & \textbf{0.1762} \\
LoCoMo & Open-domain & \textbf{0.1084} & 0.0983 & 0.1026 \\
LoCoMo & Adversarial & \textbf{0.8004} & 0.7964 & \textbf{0.8004} \\
LoCoMo & Overall & 0.5114 & 0.5021 & \textbf{0.5248} \\
\hline
\end{tabular}
}
\end{table}

As shown in Table~\ref{tab:long_context_memory_results}, graph-based methods are particularly useful when information must be aggregated across documents or sessions. On LongBench-v2, HippoRAG2 performs best on Multi-Document reasoning, while RAG achieves the highest score on Long In-context Learning. On LoCoMo, HippoRAG2 achieves the best overall performance and improves over RAG on Single-session, Multi-session, and Temporal questions. These results support our main conclusion: GraphRAG-style methods are beneficial for distributed information aggregation, while standard RAG remains strong for detail-oriented retrieval and in-context learning.

We further analyze end-to-end token usage across construction, retrieval, and generation. Construction tokens include both LLM tokens used for graph construction or summarization and embedding tokens used for indexing. Retrieval tokens refer to additional LLM calls during retrieval, when applicable. Generation tokens correspond to final answer generation.

\begin{table}[t]
\centering
\caption{End-to-end token usage on LongBench-v2 and LoCoMo.}
\label{tab:end_to_end_token_cost}
\resizebox{\linewidth}{!}{
\begin{tabular}{llccc}
\hline
Dataset & Stage & RAG & RaptorRAG & HippoRAG2 \\
\hline
LongBench-v2 & Construction-LLM & 0 & 50.2M & 184.5M \\
LongBench-v2 & Construction-Embed & 32.3M & 35.2M & 34.0M \\
LongBench-v2 & Retrieval-LLM & 0 & 0 & 1.5M \\
LongBench-v2 & Generation & 622K & 521K & 624K \\
LongBench-v2 & Total & \textbf{32.9M} & 85.9M & 220.6M \\
\hline
LoCoMo & Construction-LLM & 0 & 287.7K & 1.18M \\
LoCoMo & Construction-Embed & 218.4K & 239.3K & 222.4K \\
LoCoMo & Retrieval-LLM & 0 & 0 & 11.30M \\
LoCoMo & Generation & 4.72M & 3.65M & 4.72M \\
LoCoMo & Total & 4.94M & \textbf{4.18M} & 17.42M \\
\hline
\end{tabular}
}
\end{table}

Table~\ref{tab:end_to_end_token_cost} shows that graph-based methods often require substantially higher token budgets. RaptorRAG introduces additional construction-stage LLM cost for hierarchical summarization, while HippoRAG2 incurs high construction and retrieval-time LLM costs due to graph construction and graph-based reasoning. In contrast, standard RAG mainly uses embedding tokens during indexing and generation tokens during inference. Generation costs are relatively similar across methods under comparable retrieval settings, although RaptorRAG can use fewer generation tokens because its retrieved summaries are typically shorter than raw text chunks.

Overall, these results verify the conclusions from QA and summarization task could be generalized to other tasks.

\section{Conclusion}
\label{sec:conclusion}
This work presents a unified benchmark evaluation of RAG and GraphRAG across both question answering and query-based summarization, clarifying when explicit graph structures help, and when they do not, under controlled settings. Our analyses reveal strong task-dependent behaviors: RAG is consistently effective for single-hop, detail-oriented queries that require precise evidence, whereas GraphRAG is more advantageous for multi-hop, reasoning-intensive QA and tends to produce more corpus-level, diverse summaries.
Motivated by these findings, we study two hybrid strategies, \textbf{Selection} and \textbf{Integration}, that combine the strengths of both paradigms and improve QA performance. Beyond effectiveness, we highlight practical challenges that limit current GraphRAG systems, including incomplete or noisy graph construction, additional computation and storage overhead, and evaluation artifacts such as position effects in LLM-as-a-Judge for summarization. Together, these observations point toward the next generation of RAG systems: approaches that can construct and refine graphs reliably, adapt retrieval and aggregation to query needs, and deliver stronger reasoning benefits under realistic efficiency constraints.



\begin{acks}
Haoyu Han, Li Ma, Kai Guo, Hui Liu and Jiliang Tang are supported by the National Science Foundation (NSF) under grant numbers CNS2321416, IIS2212032, IIS2212144, IIS 2504089, DUE2234015, CNS2246050, DRL2405483 and IOS2035472, the Michigan Department of Agriculture and Rural Development, US Dept of Commerce, Gates Foundation, Amazon Faculty Award, Meta, NVIDIA, Microsoft and SNAP. Yu Wang, Yongjia Lei and Zhisheng Qi are supported by the National Science Foundation (NSF) under grant number IIS 2524379 and NAIRR 250188.
\end{acks}

\bibliographystyle{ACM-Reference-Format}
\balance
\bibliography{mybib}

\clearpage

\appendix
\section*{Appendix}
\label{sec:appendix}
\FloatBarrier

\section{Retrieval accuracy of different methods}
\label{app:sec:retrieve}

In this section, we compare the retrieval effectiveness of different methods. Since retrieval does not have explicit ground-truth supervision at the chunk level, we measure \emph{retrieval accuracy} as the proportion of examples for which the ground-truth answer string appears in the retrieved context. We report results on the \textbf{HotpotQA} and \textbf{NQ} datasets.

\begin{table}[!htb]
\centering
\caption{Retrieval accuracy (\%) of different methods on Hotpot and NQ datasets}
\label{app:tab:retrieve}
\begin{tabular}{c|cc}
\hline
Method                      & Hotpot & NQ     \\ \hline
RAG                         & 88.60  & 86.70  \\
KG-GraphRAG (Triplets only)  & 39.20  & 32.18 \\
KG-GraphRAG (Triplets+Text)  & 69.80 & 61.50 \\
Community-GraphRAG (Local)  & 67.53 & 42.20 \\
Community-GraphRAG (Global) & 88.60 & 83.30 \\ \hline

\end{tabular}
\end{table}

As shown in the Table~\ref{app:tab:retrieve}, KG-GraphRAG (Triplets only) achieves relatively low retrieval accuracy, particularly on NQ. This is primarily due to the incompleteness of the constructed knowledge graphs—only 65.8\% of answer entities exist in the HotpotQA KG, and 65.5\% in the NQ KG. In contrast, Community-GraphRAG, which leverages community-level summarization, demonstrates significantly better retrieval performance.
These findings highlight several potential directions for improvement:
\begin{enumerate}[leftmargin=1em]
\item Enhancing KG construction to increase entity and relation coverage.
\item Combining structured graph information with raw text to improve retrieval robustness and completeness.
\end{enumerate}

\section{Graph Construction with different LLMs}
\label{app:graphconstruction}
In the main paper, we use GPT-4o-mini to extract entities and relationships for graph construction due to cost considerations. To investigate whether stronger LLMs yield better performance, we also use GPT-4o for graph extraction. Specifically, we evaluate this on the MultiHop-RAG and ODSum-story datasets, representing question answering and summarization tasks, respectively.  We focused on Community-GraphRAG (Local) as a representative method (GraphRAG) and evaluated it with both LLaMA3.1-8B and LLaMA3.1-70B for generation. 

The results are shown in Table~\ref{app:tab:cons1}, Table~\ref{app:tab:cons2}, Table~\ref{app:tab:cons3} and Table~\ref{app:tab:cons4}, respectively. The results show that using a stronger LLM (GPT-4o) for graph extraction generally improves the performance of GraphRAG on both question answering and summarization tasks. However, the overall conclusion regarding the relative performance of RAG and GraphRAG remains consistent across different graph construction backbones.

\begin{table}[H]
\centering
\caption{Performance of different graph construction methods with Llama 3.1–8B on the MultiHop-RAG dataset.}
\label{app:tab:cons1}
\resizebox{0.45\textwidth}{!}{%
\begin{tabular}{c|ccccc}
\hline
          & Inference & Comparison & Null  & Temporal & Overall \\ \hline
RAG         & 92.16     & 57.59       & 96.01 & 30.7     & 67.02   \\
GPT-4o-mini & 86.89     & 60.63       & 80.07 & 50.6     & 69.01   \\
GPT-4o      & 88.11     & 62.62       & 70.43 & 49.74    & 68.74   \\ \hline
\end{tabular}
}
\end{table}

\begin{table}[H]
\centering
\caption{Performance of different graph construction methods with Llama 3.1–70B on the MultiHop-RAG dataset.}
\label{app:tab:cons2}
\resizebox{0.45\textwidth}{!}{%
\begin{tabular}{c|ccccc}
\hline
70B         & Inference & Comparison & Null  & Temporal & Overall \\ \hline
RAG         & 94.85     & 56.31       & 91.36 & 25.73    & 65.77   \\
GPT-4o-mini & 92.03     & 60.16       & 88.70  & 49.06    & 71.17   \\
GPT-4o      & 93.63     & 66.59       & 81.06 & 58.49    & 75.08   \\ \hline
\end{tabular}
}
\end{table}

\begin{table}[H]
\centering
\caption{Performance of different graph construction methods with Llama 3.1–8B on the ODSum-story dataset.}
\label{app:tab:cons3}
\begin{tabular}{c|ccc|ccc}
\hline
          & \multicolumn{3}{c|}{ROUGE-2} & \multicolumn{3}{c}{BERTScore} \\ \hline
            & P        & R       & F1      & P        & R        & F1      \\ \hline
RAG         & 15.39    & 8.44    & 9.81    & 83.87    & 85.74    & 84.57   \\
GPT-4o-mini & 13.84    & 7.19    & 8.49    & 83.19    & 85.07    & 83.90    \\
GPT-4o      & 13.99    & 7.45    & 8.64    & 83.24    & 85.1     & 83.94   \\ \hline
\end{tabular}
\end{table}

\begin{table}[H]
\centering
\caption{Performance of different graph construction methods with Llama 3.1–8B on the ODSum-meeting dataset.}
\label{app:tab:cons4}
\begin{tabular}{c|ccc|ccc}
\hline
                   & \multicolumn{3}{c|}{ROUGE-2} & \multicolumn{3}{c}{BERTScore} \\ \hline
\multicolumn{1}{l|}{} & P        & R       & F1      & P        & R        & F1      \\ \hline
RAG                   & 11.85    & 14.24   & 11.09   & 85.96    & 85.76    & 85.67   \\
GPT-4o-mini           & 12.54    & 10.31   & 9.61    & 84.51    & 85.33    & 84.71   \\
GPT-4o                & 12.08    & 10.84   & 9.72    & 84.66    & 85.28    & 84.77   \\ \hline
\end{tabular}
\end{table}

\section{Token-Matched Retrieval Analysis}
\label{app:sec:token_matched}

Beyond runtime and storage, we further analyze whether performance differences are affected by the number of retrieved input tokens. In our default setting, RAG retrieves the top-10 text chunks, while Community-GraphRAG retrieves the top-10 entities together with their associated entity descriptions, relations, relation descriptions, and community summaries. As a result, Community-GraphRAG uses substantially more retrieved tokens than RAG: 9,770 vs. 3,631 tokens on MultiHop-RAG, and 10,244 vs. 2,279 tokens on ODSum-Story.

To control for this difference, we conduct a token-matched experiment by increasing the number of retrieved chunks for RAG so that its input length is comparable to Community-GraphRAG. We report the results using Llama~3.1--70B in Tables~\ref{app:tab:token_multihop} and~\ref{app:tab:token_odsum}.

\begin{table*}[!htb]
\centering
\caption{Performance comparison of RAG, token-matched RAG, and GraphRAG using Llama~3.1--70B on MultiHop-RAG.}
\label{app:tab:token_multihop}
\small
\setlength{\tabcolsep}{6pt}
\begin{tabular}{lccccc}
\hline
Method & Inference & Comparison & Null & Temporal & Overall \\
\hline
RAG & 94.85 & 56.31 & 91.36 & 25.73 & 65.77 \\
RAG-SameToken & 95.96 & 59.58 & 88.70 & 43.74 & 71.01 \\
GraphRAG & 92.03 & 60.16 & 88.70 & 49.06 & 71.17 \\
\hline
\end{tabular}
\end{table*}

\begin{table*}[!htb]
\centering
\caption{Performance comparison of RAG, token-matched RAG, and GraphRAG using Llama~3.1--70B on ODSum-Meeting.}
\label{app:tab:token_odsum}
\small
\setlength{\tabcolsep}{6pt}
\begin{tabular}{lccc|ccc}
\hline
Method & \multicolumn{3}{c|}{ROUGE-2} & \multicolumn{3}{c}{BERTScore} \\
\cline{2-7}
 & P & R & F1 & P & R & F1 \\
\hline
RAG & 11.85 & 14.24 & 11.09 & 85.96 & 85.76 & 85.67 \\
RAG-SameToken & 12.82 & 14.07 & 11.34 & 85.86 & 86.00 & 85.73 \\
GraphRAG & 12.54 & 10.31 & 9.61 & 84.51 & 85.33 & 84.71 \\
\hline
\end{tabular}
\end{table*}

The results show that increasing RAG's token budget improves performance, especially on MultiHop-RAG. However, the main conclusions remain unchanged. RAG and token-matched RAG remain strong on inference-style queries and summarization tasks, where detailed evidence can be directly retrieved from text chunks. In contrast, GraphRAG remains competitive or better on Comparison and Temporal questions, which require multi-hop reasoning and information aggregation. These results suggest that the differences between RAG and GraphRAG cannot be explained solely by retrieved input length.

\section{Additional Analysis of Position Bias in LLM-as-a-Judge Evaluation}
\label{app:position_bias}

\begin{table*}[!htb]
\centering
\caption{Additional LLM-as-a-Judge results for GraphRAG-Local (GR-Local) vs. RAG. We report the winning ratios under two presentation orders, together with the order-averaged GraphRAG-Local win rate.}
\label{tab:app_position_local}
\begin{tabular}{lllccc}
\hline
Dataset & Judge & Metric & O1 (GR-Local:RAG) & O2 (RAG:GR-Local) & Mean GR-Local \\
\hline
QMSum & GPT-4o & Comp & 0.97:0.03 & 0.88:0.12 & 0.54 \\
QMSum & GPT-4o & Div  & 0.91:0.09 & 0.18:0.82 & 0.87 \\
QMSum & GPT-5.4 & Comp & 0.44:0.56 & 0.88:0.12 & 0.28 \\
QMSum & GPT-5.4 & Div  & 0.62:0.38 & 0.62:0.38 & 0.50 \\
\hline
ODSum-story & GPT-4o & Comp & 0.62:0.38 & 0.74:0.26 & 0.44 \\
ODSum-story & GPT-4o & Div  & 0.63:0.37 & 0.45:0.55 & 0.59 \\
ODSum-story & GPT-5.4 & Comp & 0.47:0.53 & 0.77:0.23 & 0.35 \\
ODSum-story & GPT-5.4 & Div  & 0.56:0.44 & 0.56:0.44 & 0.50 \\
\hline
\end{tabular}
\end{table*}

\begin{table*}[!htb]
\centering
\caption{Additional LLM-as-a-Judge results for GraphRAG-Global (GR-Global) vs. RAG. We report the winning ratios under two presentation orders, together with the order-averaged GraphRAG-Global win rate.}
\label{tab:app_position_global}
\begin{tabular}{lllccc}
\hline
Dataset & Judge & Metric & O1 (GR-Global:RAG) & O2 (RAG:GR-Global) & Mean GR-Global \\
\hline
QMSum & GPT-4o & Comp & 0.59:0.41 & 1.00:0.00 & 0.29 \\
QMSum & GPT-4o & Div  & 0.91:0.09 & 0.09:0.91 & 0.91 \\
QMSum & GPT-5.4 & Comp & 0.06:0.94 & 1.00:0.00 & 0.03 \\
QMSum & GPT-5.4 & Div  & 0.59:0.41 & 0.44:0.56 & 0.57 \\
\hline
ODSum-story & GPT-4o & Comp & 0.26:0.74 & 0.94:0.06 & 0.16 \\
ODSum-story & GPT-4o & Div  & 0.85:0.15 & 0.26:0.74 & 0.79 \\
ODSum-story & GPT-5.4 & Comp & 0.12:0.88 & 0.97:0.03 & 0.08 \\
ODSum-story & GPT-5.4 & Div  & 0.61:0.39 & 0.53:0.47 & 0.54 \\
\hline
\end{tabular}
\end{table*}

To assess the position bias, we conduct additional pairwise evaluations between RAG and GraphRAG (GR) on QMSum and ODSum-story. We consider two mitigation strategies: (i) swapping the presentation order and reporting the averaged win rate, and (ii) using multiple judges, including GPT-4o and GPT-5.4, to examine whether the conclusions are robust across judge models.

Tables~\ref{tab:app_position_local} and~\ref{tab:app_position_global} report the results for GraphRAG-Local and GraphRAG-Global, respectively. For each setting, we show the winning ratios under two orders, O1 and O2, for Comprehensiveness (Comp) and Diversity (Div). The last column reports the order-averaged GraphRAG win rate, where the GraphRAG score is taken from the corresponding side of each ratio.

We observe that position bias exists for both judges, but its magnitude varies by model. GPT-4o is more order-sensitive, often strongly favoring the first-presented summary. GPT-5.4 reduces this effect but does not fully eliminate it. Despite this bias, the main trends remain stable after order averaging. GraphRAG-Global consistently improves Diversity, confirming that global community-level summaries better capture broad thematic coverage. In contrast, RAG performs better on Comprehensiveness, likely because direct retrieval preserves more fine-grained, query-specific evidence. GraphRAG-Local is generally competitive with RAG, especially on Diversity, where its averaged win rates are close to balanced under GPT-5.4.

These results suggest a simple evaluation protocol for reducing position bias: evaluate both answer orders and report the averaged win rate. When possible, using multiple judges further helps verify the robustness of the conclusions.

\end{document}